\documentclass[fleqn,usenatbib]{mnras}

\usepackage{newtxtext,newtxmath}
\usepackage[T1]{fontenc}
\DeclareRobustCommand{\VAN}[3]{#2}
\let\VANthebibliography\thebibliography
\def\thebibliography{\DeclareRobustCommand{\VAN}[3]{##3}\VANthebibliography}

\usepackage{graphicx}	% Including figure files
\usepackage{amsmath}	% Advanced maths commands
\usepackage{placeins}
\usepackage{caption}
\usepackage{afterpage}

\title[LIV and KM3NeT Cosmogenic Neutrino]{Probing Lorentz Invariance Violation in Cosmogenic Neutrino Propagation with KM3-230213A}

\author[Rodrigo Sasse, Rodrigo Guedes Lang and Rita C. Anjos]{
Rodrigo Sasse,$^{1,2}$\thanks{E-mail: rodrigo.sasse1@uel.br}
Rodrigo Guedes Lang,$^{2}$\thanks{E-mail: rodrigo.lang@fau.de}
Rita C. Anjos,$^{1,3,4,5,6}$\thanks{E-mail: ritacassia@ufscar.br}
\\
$^{1}$Programa de pós-graduação em Física \& Departamento de Física, Universidade Estadual de Londrina (UEL)\\ Rodovia Celso Garcia Cid Km 380, 86057-970 Londrina, PR, Brazil\\
$^{2}$Friedrich-Alexander-Universität Erlangen-Nürnberg, Erlangen Centre for Astroparticle Physics, Nikolaus-Fiebiger-Str. 2, 91058 Erlangen, Germany\\
$^{3}$Centro de Artes, Humanidades e Tecnologia, Universidade Federal de São Carlos (UFSCar), R. Dr. Eduardo Nielsen, 420, Jardim Congonhas\\, 15030-070 São José do Rio Preto, SP, Brazil\\
$^{4}$Programa de Pós-Graduação em Física e Astronomia, Universidade Tecnológica Federal do Paraná, 80230-901 Curitiba, PR , Brazil\\
$^{5}$N\'ucleo de Astrof\'{\i}sica e Cosmologia, Universidade Federal do Esp\'irito Santo, 29075-910 Vit\'oria, ES, Brazil\\
$^{6}$Programa de Pós-Graduação em Física Aplicada, Universidade Federal da Integração Latino-Americana, 85867-670 Foz do Igua\c{c}u, PR, Brazil
}

\date{Accepted XXX. Received YYY; in original form ZZZ}

\pubyear{\the\year{}}

\begin{document}
\label{firstpage}
\pagerange{\pageref{firstpage}--\pageref{lastpage}}
\maketitle

\begin{abstract}
We investigate superluminal Lorentz invariance violation (LIV) in the neutrino sector using cosmogenic neutrino fluxes generated with ultrahigh-energy cosmic-ray propagation models. Standard fluxes are calculated with \texttt{CRPropa 3.2} and subsequently modified using a prescription based on LIV-induced neutrino splitting. Superluminal LIV suppresses the flux at the highest energies while producing an enhancement at PeV - EeV energies. We use the KM3-230213A event as a benchmark to evaluate the sensitivity of current observations to these spectral modifications. Although the available statistics do not allow a formal constraint, the predicted fluxes are particularly sensitive to coefficients in the range
$10^{-24}\,\mathrm{eV}^{-1} \lesssim \delta_{\nu,1} \lesssim 10^{-22}\,\mathrm{eV}^{-1}$,
with the results strongly depending on the assumed cosmic-ray source properties. Intermediate coefficients can enhance the expected event rate within the reconstructed energy range of KM3-230213A, whereas larger coefficients may overproduce neutrinos in energy intervals constrained by IceCube and the Pierre Auger Observatory. These results identify a region of observational sensitivity to LIV and provide testable predictions for future neutrino telescopes.
\end{abstract}


%%%%%%%%%%%%%%%%%%%%%%%%%%%%%%%%%%%%%%%%%%%%%%%%%%

%%%%%%%%%%%%%%%%% BODY OF PAPER %%%%%%%%%%%%%%%%%%

\section{Introduction}

The recent detection of an ultra-high-energy (UHE) neutrino event at an estimated energy of 220$^{+570}_{-110}$ PeV by the KM3NeT collaboration (KM3-230213A) has opened a new observational window on the highest-energy neutrino sky and provides a unique probe of fundamental physics~\citep{2025Natur.638..376K, KM3NeT2025Cosmogenic}. Its origin, whether atmospheric, astrophysical, or cosmogenic, is actively debated~\citep{KM3NeT2025GlobalLandscape, 2025arXiv250208484K, KM3NeT2025Galactic, KM3NeT2025GRB, Adriani2025, 2025PhRvD.112f3045B, 2025PhRvL.135l1003K, 2025PhRvD.111l3022B, 2025arXiv251026126S, Satunin2025, 2025PhRvD.112h3061S, yang2025}. If produced via photopion interactions of ultra-high-energy cosmic rays (UHECRs) with background photons ($p + \gamma_{\rm bg} \to \Delta^+ \to p/n + \pi^{0/+}$; $\pi^\pm \to \mu^\pm \nu_\mu(\bar{\nu}_\mu) \to e^\pm \nu_e(\bar{\nu}_e) \nu_\mu \bar{\nu}_\mu$), the event lies in tension with the cosmogenic neutrino fluxes expected from UHECR models that successfully describe the observed spectrum and composition~\citep{2025ApJ...994...31S, ANCHORDOQUI20191}, as well as with current upper limits on the diffuse flux established by the Pierre Auger Observatory~\citep{AbdulHalim:2023SN} and IceCube~\citep{PhysRevD.98.062003}.

Regardless of the ultimate origin of KM3-230213A, the event represents an important observational mechanism for testing non-standard neutrino physics. One well-motivated candidate is Lorentz invariance violation (LIV). Lorentz invariance (LI), a cornerstone of the Standard Model, is expected to break near the Planck scale in various quantum-gravity frameworks~\citep{PhysRevD.58.116002, ADDAZI2022103948}. Ultra-high-energy astroparticles are exceptionally sensitive probes of such effects due to their extreme energies and cosmological propagation distances~\citep{2018ApJ...853...23G, Abreu_2022, Lang2022}. LIV is typically introduced via a modified dispersion relation (MDR), which leads to energy-dependent particle velocities that deviate from the speed of light, resulting in anomalous time-of-flight measurements and altered kinematic thresholds. In the superluminal case ($v > c$), neutrinos become unstable and open decay channels kinematically forbidden in the Standard Model, substantially modifying the observed flux at Earth~\citep{Carmona_2026}.

Studies of LIV with UHE astroparticles have largely focused on the electromagnetic and hadronic sectors, examining modifications to interaction thresholds (e.g., pair production, photopion interactions) and their impact on observed spectra and time delays~\citep{2018ApJ...853...23G, Lang2022, 2025EPJC...85..604L}. In the neutrino sector, previous works have primarily investigated superluminal effects such as vacuum pair emission, neutrino splitting, and time-of-flight anomalies~\citep{2011PhRvL.107r1803C, PhysRevD.107.043001, Carmona_2024, Carmona_2026}. Crucially, cosmogenic neutrinos propagate without further interactions, simplifying LIV analysis compared to the complex cascade calculations required for UHE photons. Superluminal LIV effects on the neutrino spectrum can be incorporated through an analytical transformation of the standard flux, making the calculation both tractable and robust.

In this work, we characterise superluminal LIV in the neutrino sector through its effect on the predicted cosmogenic neutrino flux, using KM3-230213A as a benchmark to probe the LIV parameter space. We implement LIV-modified neutrino propagation within the \texttt{CRPropa 3.2}~\citep{2022JCAP...09..035A} toolkit and map out how the resulting spectral modifications depend on the LIV coefficient and the UHECR source model. Section~\ref{sec:liv} reviews the LIV framework and modified dispersion relations; Section~\ref{sec:method} describes our methodology; Section~\ref{sec:results} presents the results and their implications; and Section~\ref{sec:conclusions} provides a summary and outlook.

\section{Lorentz Invariance Violation Framework}
\label{sec:liv}

The standard framework for testing LIV in astroparticle physics employs a MDR that introduces energy-dependent corrections to the particle's energy-momentum relation~\citep{Lang2022,2022PrPNP.12503948A,2020Symm...12.1232M}. For a particle of species $a$ with mass $m_a$, the MDR can be written as
\begin{equation}
E_a^2 = m_a^2 + p_a^2 +\sum_{n=0,1,2,\ldots} \delta_{a,n} E_a^{n+2},
\label{eq:MDR}
\end{equation}
where $E_a$ and $p_a$ are the particle's energy and momentum, and $\delta_{a,n}$ are the LIV parameters with dimensions of $\text{E}^{-n}$. The summation index $n$ denotes the order of the LIV term; commonly considered are the linear ($n=1$) and quadratic ($n=2$) corrections, although a constant shift ($n=0$) can also lead to interesting phenomenological effects. The sign of $\delta_{a,n}$ determines whether the particle becomes superluminal ($\delta_{a,n}>0$) or subluminal ($\delta_{a,n}<0$) at high energies. For $n>0$, the LIV scale is conventionally defined as $E_{\text{LIV}}^{(n)} = |\delta_{a,n}|^{-1/n}$, which in quantum-gravity inspired scenarios is often associated with the Planck scale ($E_{\text{Pl}} \approx 1.22\times10^{19}$~GeV) or other fundamental scales~\citep{1998Natur.393..763A, 2006AnPhy.321..150J}.

For photons ($m_\gamma=0$), the MDR simplifies to
\begin{equation}
E_\gamma^2 = p_\gamma^2 + \sum_{n=0,1,2,\ldots} \delta_{\gamma,n} E_\gamma^{n+2},
\label{eq:MDR_photon}
\end{equation}

%\begin{equation}
%E_\gamma^2 = p_\gamma^2\Bigg[1 + \sum_{n} \delta_{\gamma,n} E_\gamma^n\Bigg],
%\label{eq:MDR_photon}
%\end{equation}
\noindent which modifies the kinematics of pair production ($\gamma\gamma_{\text{bg}} \to e^+e^-$) and can lead to either increased or decreased absorption depending on the sign of $\delta_{\gamma,n}$~\citep{2001APh....16...97S, 2008PhRvL.100b1102G}. Similarly, for ultra-high-energy cosmic rays (UHECRs), LIV in the hadronic sector alters the thresholds for photopion production and photodisintegration, affecting the expected spectrum and composition at Earth~\citep{2009APh....31..220S,2009PhRvD..79h3015B, Abreu_2022,Lang_2024}.

In the neutrino sector, which is the focus of this work, the nearly massless nature of neutrinos ($m_\nu \ll E_\nu$) leads to the simplified MDR:
$E_\nu^2 \approx p_\nu^2 + \sum_{n} \delta_{\nu,n} E_\nu^{n+2}$. Such modifications can induce decay processes such as vacuum pair emission ($\nu \to \nu e^+e^-$) or neutrino splitting ($\nu \to \nu\nu\bar{\nu}$) for superluminal neutrinos, reducing their survival probability over cosmological distances~\citep{2011PhRvL.107r1803C,Carmona_2026}, while subluminal neutrinos experience an effective reduction in propagation speed that can enhance their survival by suppressing certain interaction channels.

In this work we consider superluminal LIV ($\delta_{\nu,1}>0$) for cosmogenic neutrinos. Unlike UHE photons or cosmic rays, neutrinos propagate without further interactions, allowing the effects of LIV to be incorporated via a simple analytic transformation of the standard spectrum. This approach avoids the complexities of cascade calculations and provides a tractable framework for probing LIV with the cosmogenic neutrino flux, using KM3-230213A as a reference event.

\section{Simulation Framework and LIV Implementation}
\label{sec:method}

\subsection{UHECR Sources and Cosmogenic Neutrino Production}

The propagation of UHECRs and the self-consistent production of secondary cosmogenic neutrinos were simulated using the \texttt{CRPropa 3.2} framework. We adopted a one-dimensional approximation to simulate the propagation of five representative primary nuclei: Hydrogen ($^1$H), Helium ($^4$He), Nitrogen ($^{14}$N), Silicon ($^{28}$Si), and Iron ($^{56}$Fe). Their abundances were weighted to match the Pierre Auger Observatory best-fit source composition parameters. In this model, the abundances are defined as differential flux fractions at a reference energy of $10^{18}$ eV, dominated by $\approx 67.3\%$ Helium, Nitrogen $\approx 28.1\%$ and Silicon $\approx 4.6\%$ at source injection~\citep{Aab_2017}. The simulation covered an energy range from $0.5 \times 10^{18}$~eV to $10^{21}$~eV over a comoving distance up to $D_{\rm max} = 8000$~Mpc, ensuring the inclusion of sources up to redshift $z \approx 5$.

A critical component of our model is the source density evolution, parameterized as $\Psi(z) \propto (1+z)^m$, where $m$ controls the cosmological evolution rate. We test scenarios with $m \in [0.0, 2.0]$, representing conservative source evolution models, given that Pierre Auger Observatory results do not support strong source evolution rates \citep{Batista_2019}. This parameter affects the normalization and shape of the injected spectrum but is treated within the standard astrophysical framework. Systematic uncertainties in these source models constitute the dominant uncertainty in our predictions.

We adopted two models to describe the observed spectrum and composition. The first follows the combined fit parameters based on the Pierre Auger Observatory data~\citep{Aab_2017}. This component is characterized by a hard injection spectrum ($\alpha \approx 0.96$), a low rigidity cutoff ($R_{\mathrm{max}} = 4.5 \times 10^{18}$ eV) and the best-fit composition dominated by intermediate-mass nuclei. In the second, a two-component scenario, we combined the best fit parameters with a pure proton composition that was added accounting for 10\% of the total injected composition at the sources featuring a softer spectrum ($\alpha_p = 1.0 ~-~2.5$) and a high rigidity cutoff ($R_{\mathrm{max}} = 4.5 \times 10^{18}$ to $1 \times 10^{21}$~eV)\citep{PhysRevD.100.103008}.~Finally, the overall flux normalization is fixed by matching the total all-particle cosmic-ray spectrum measured by the Pierre Auger Collaboration~\citep{p4l5-hxlf} at a reference energy of $E_{\rm norm} = 10^{18.75}$~eV.

%. For this model, we investigate the impact of the source evolution by varying the parameter $m$

 %The second is a sub-dominant light component consisting of pure protons with a softer spectrum ($\alpha = 2.0$) and a high rigidity cutoff ($R_{\rm max} = 5 \times 10^{20}$~eV). For this case, we assumed a strong source evolution with $m=5.0$. The total flux is constructed as a weighted sum of these components, with a 10\% contribution from the high-energy proton component added to the mixed-composition baseline. Finally, the overall flux normalization is fixed by matching the total all-particle cosmic-ray spectrum measured by the Pierre Auger Collaboration~\citep{Aab_2017} at a reference energy of $E_{\rm norm} = 10^{18.75}$~eV.

Relevant energy-loss mechanisms that have been considered include photopion production ($p + \gamma_{\rm bg} \to \Delta^+ \to p/n + \pi^{0/+}$), photodisintegration via $A + \gamma_{\rm bg} \to (A-1) + N$, and Bethe-Heitler pair production. The interaction rates are computed using detailed models of the dominant photon backgrounds, specifically the Cosmic Microwave Background (CMB) and the Extragalactic Background Light (EBL). The CMB is treated as a blackbody with temperature $T(z) = 2.73(1 + z)$ K, while the EBL is modeled following the prescription from \citet{10.1111/j.1365-2966.2012.20841.x}. The resulting neutrino flux, obtained by recording the properties of neutrinos produced via pion and muon decays, serves as the input for our LIV study.

\subsection{Neutrino Propagation with LIV}

To simulate the superluminal scenario described in Sec.~\ref{sec:liv}, we implement the decay rates for the dominant channels. Specifically, we focus on neutrino splitting ($\nu \to 3\nu$) decay channel, as this acts as the dominant mechanism for the spectral cutoff and the resulting flux regeneration. The decay rate is highly sensitive to energy, scaling as $\Gamma_{\rm spl} \propto E^{5 + 3n}$~\citep{PhysRevD.107.043001, Carmona_2024}. Consequently, the survival probability $P_{\rm surv}$ for a neutrino with energy $E$ traveling a comoving distance $D(z)$ is determined by the accumulated decay rate, computed formally as:

\begin{equation}
P_{\rm surv}(E, z) = \exp\left( - \int_{0}^{z} \frac{\Gamma_{\rm spl}(E(z'))}{H(z')(1+z')} \, dz' \right),
\label{eq:prob_surv}
\end{equation}

\noindent where $H(z')$ represents the Hubble parameter at redshift $z'$ and the factor $(1+z')^{-1}$ accounts for cosmological time dilation along the trajectory~\citep{Carmona_2026}. To incorporate the pile-up effect, we implement a reweighting scheme. Since the splitting process produces three daughter neutrinos, each carrying approximately a fraction $f \approx 1/3$ of the parent energy, the modified differential flux $\Phi_{\rm LIV}(E)$ is constructed as the superposition of the surviving primary flux and the redistributed secondary flux:

\begin{equation}
\begin{split}
\Phi_{\rm LIV}(E) &= \Phi_{\rm std}(E) P_{\rm surv}(E) \\
&\quad + 3 \int_{E}^{\infty} \Phi_{\rm std}(E') [1 - P_{\rm surv}(E')] \delta(E - f E') \, dE'.
\end{split}
\label{eq:pileup}
\end{equation}

\noindent Here, $\Phi_{\rm std}(E)$ denotes the standard neutrino flux expected under LI assumptions. The second term describes the contribution of decayed neutrinos, where integration is performed on all possible parent energies $E'$. The Dirac delta function $\delta(E - f E')$ enforces the specific kinematic condition where secondary neutrinos inherit a fixed energy fraction $f$ of the parent. This approach ensures particle number conservation while accurately modeling the spectral suppression at high energies and the consequent enhancement at lower energies.

\begin{figure}
\centering
\includegraphics[width=0.99\columnwidth]{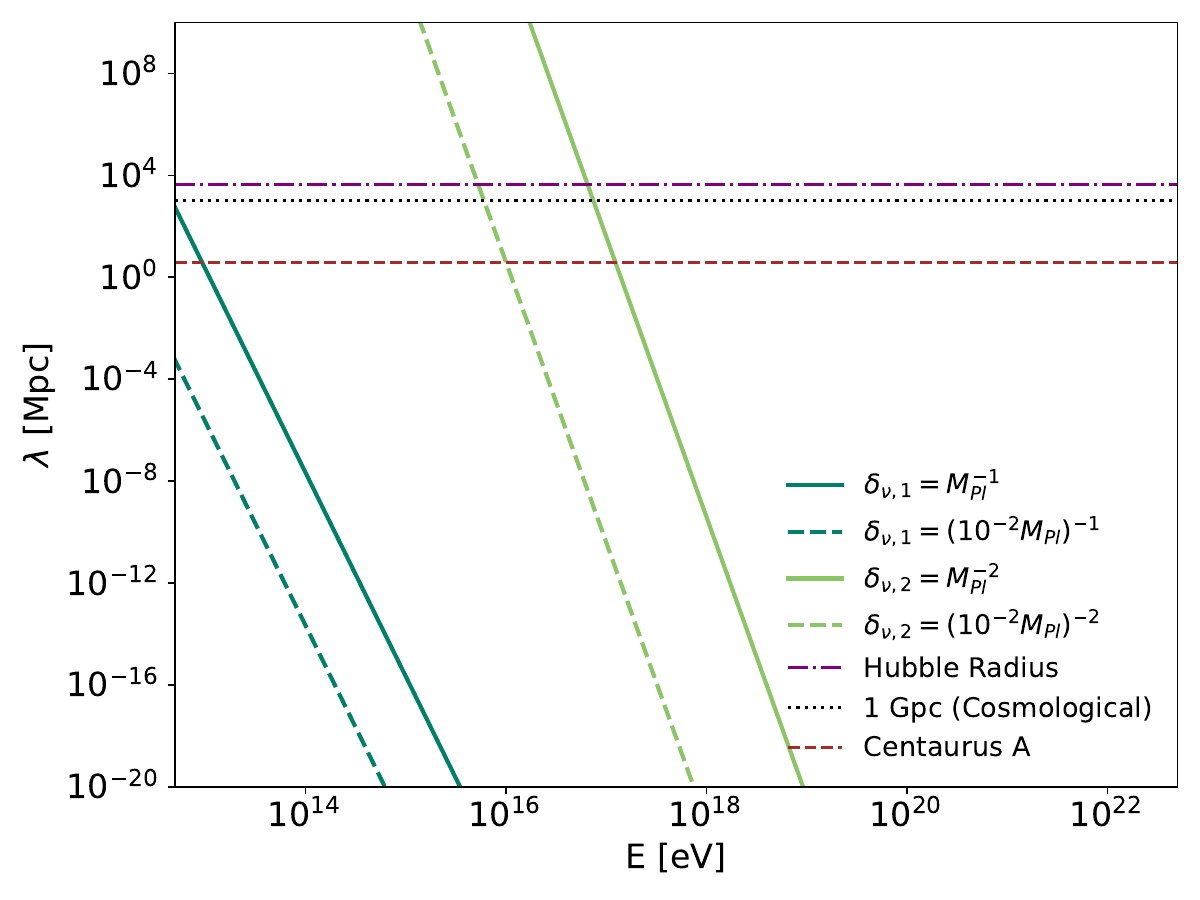}
%\caption{Neutrino mean free path $\lambda_{\rm MFP}$ as a function of energy for superluminal LIV scenarios. The solid curves correspond to different orders of Lorentz violation ($n=1, 2$) and the dashed lines quantum gravity scales ($\delta_{\nu,1} = M_{\rm Pl}^{-1}$ and $\delta_{\nu,2} = 10^{-2} M_{\rm Pl}$). Horizontal lines indicate reference astrophysical distances: the Hubble radius (solid line), 1 Gpc (dotted line), and the distance to Centaurus A (dashed line).}
\caption{Neutrino mean free path $\lambda_{\rm MFP}$ as a function of energy for superluminal LIV scenarios. The solid curves correspond to different orders of Lorentz violation ($n=1, 2$) and the dashed lines represent quantum gravity scales ($\delta_{\nu,1} = M_{\rm Pl}^{-1}$ and $\delta_{\nu,2} = (10^{-2} M_{\rm Pl})^{-2}$). Horizontal lines indicate reference astrophysical distances: the Hubble radius (solid line), 1 Gpc (dotted line), and the distance to Centaurus A (dashed line).}
\label{fig:mfp}
\end{figure}

To illustrate the impact of the splitting process, Figure \ref{fig:mfp} shows the neutrino mean free path, $\lambda_{\rm MFP} = c/\Gamma_{\rm spl}$, as a function of energy for different LIV orders ($n$) and energy scales ($\delta_{\nu,n}$ with n = 1 and n = 2). Whereas neutrinos are stable over cosmological distances in the Standard Model, LIV-induced decay leads to a characteristic suppression of the flux at the highest energies and a slight enhancement at lower energies. When the neutrino mean free path falls below a given propagation distance, the Universe becomes effectively opaque to neutrinos of that energy. The horizontal lines indicate representative propagation distances: the Hubble radius, 1~Gpc, and the distance to Centaurus~A.

\section{Results and Discussion}
\label{sec:results}

\subsection{Cosmogenic Neutrino Flux and LIV Parameter Space}

%We computed the expected cosmogenic neutrino flux, incorporating LIV effects for several UHECR source models that successfully describe current Auger data~\citep{Aab_2017, 2025ApJ...994...31S}. Figure~\ref{fig:exclusion} shows our exploration of the UHECR source parameter space, varying both the spectral index $\alpha_p$ and the maximum rigidity $R_{\mathrm{max,p}}$ for different LIV coefficients $\delta_{\nu,1}$ and composition scenarios characterized by the parameter $m$ ranging from 0.0 to 2.0. The color scale indicates the flux ratio limit, defined by the comparison between the predicted flux and observational sensitivities in the UHE regime ($10^{18.5}-10^{21}$)~eV. Darker regions corresponding to parameter combinations that predict fluxes below current observational constraints and brighter regions showing where predicted fluxes approach or exceed limits.

\begin{figure*}
\centering
\includegraphics[width=0.85\textwidth]{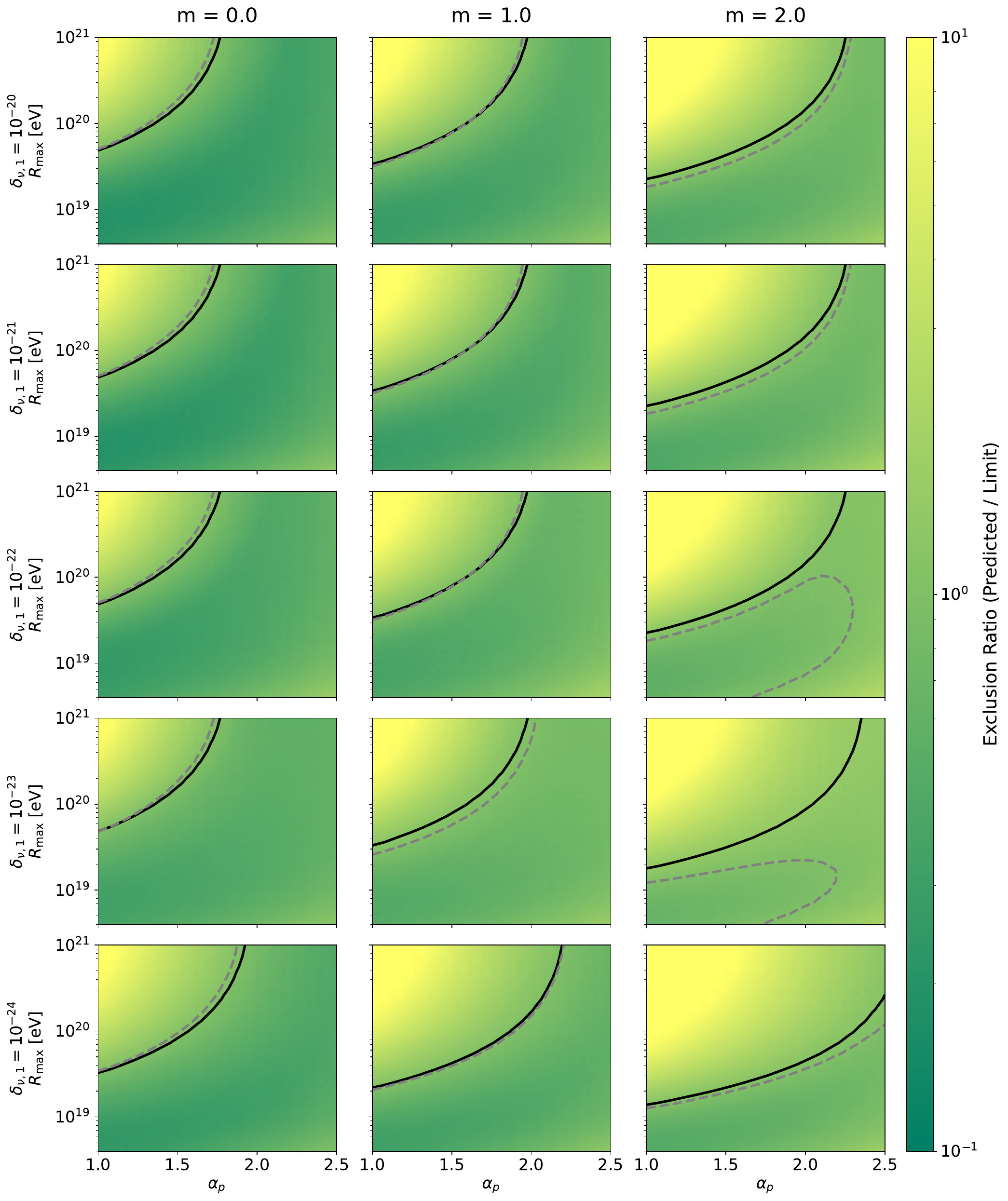}
\caption{Exclusion maps in the UHECR source parameter space of spectral index $\alpha_p$ versus maximum rigidity $R_{\mathrm{max,p}}$ for different LIV coefficients $\delta_{\nu,1}$ ranging from $10^{-20} \ \mathrm{eV}^{-1}$ to $10^{-24} \ \mathrm{eV}^{-1}$ and different composition scenarios characterized by $m = 0.0, 1.0, 2.0$. The color scale indicates the flux ratio limit, the solid black line represents the Pierre Auger Observatory upper limit and the gray dashed is the IceCube limit. Darker regions correspond to conservative models with lower predicted fluxes, while brighter regions approach observational limits.}
\label{fig:exclusion}
\end{figure*}

We computed the expected cosmogenic neutrino flux, incorporating LIV effects for several UHECR source models that successfully describe current Pierre Auger Observatory data~\citep{Aab_2017, 2025ApJ...994...31S}. Figure~\ref{fig:exclusion} shows our exploration of the UHECR source parameter space, varying both the spectral index $\alpha_p$ and the maximum rigidity $R_{\mathrm{max,p}}$ for different LIV coefficients $\delta_{\nu,1}$ and composition scenarios characterized by the parameter $m$ ranging from 0.0 to 2.0. The color scale indicates the exclusion ratio, defined as $\mathcal{R} = \Phi_{\mathrm{predicted}} / \Phi_{\mathrm{limit}}$, evaluated at the most constraining energy bin. In these maps, brighter regions (yellow, $\mathcal{R} > 1$) correspond to models that overproduce neutrinos and are thus excluded by current observations, whereas darker regions (green, $\mathcal{R} \le 1$) indicate allowed parameter space. The boundaries where $\mathcal{R} = 1$ are delineated by solid black and dashed gray contours, representing the limits set by the Pierre Auger Observatory and IceCube, respectively.

For each composition scenario, we observe a characteristic boundary in the $(\alpha_p, R_{\mathrm{max,p}})$ parameter space beyond which the predicted cosmogenic neutrino flux becomes inconsistent with previous non-detections. The position and shape of these exclusion boundaries sensitively depend on both the spectral index and maximum rigidity. Models with harder spectra corresponding to smaller values of $\alpha_p$ near 1.0 tend to produce more high-energy particles and consequently more cosmogenic neutrinos~\citep{ANCHORDOQUI20191}. Similarly, higher maximum rigidities extending beyond $10^{20}$ eV allow cosmic rays to reach the energies necessary for efficient photopion production, substantially increasing the neutrino yield~\citep{2009APh....31..220S}.

The variation across different source evolution scenarios, parametrized by the index $m$ from $m = 0.0$ to $m = 2.0$, reveals important physics. Stronger evolution scenarios are represented by higher values of $m$ and imply a larger population of active sources at high redshifts. These earlier UHECR interactions significantly enhance the production of cosmogenic neutrinos, leading to tighter constraints on the allowed parameter space. In contrast, weaker evolution with lower values of $m$ results in a reduced cosmogenic flux, thereby relaxing the constraints~\citep{2008JCAP...10..033A}. This dependence highlights the challenge in using cosmogenic neutrino predictions to constrain LIV.

%with black contours marking the boundary where predictions become inconsistent with observations
\subsection{Impact of LIV on the Cosmogenic Neutrino Spectrum}

\begin{figure*}
\centering
\includegraphics[width=\textwidth]{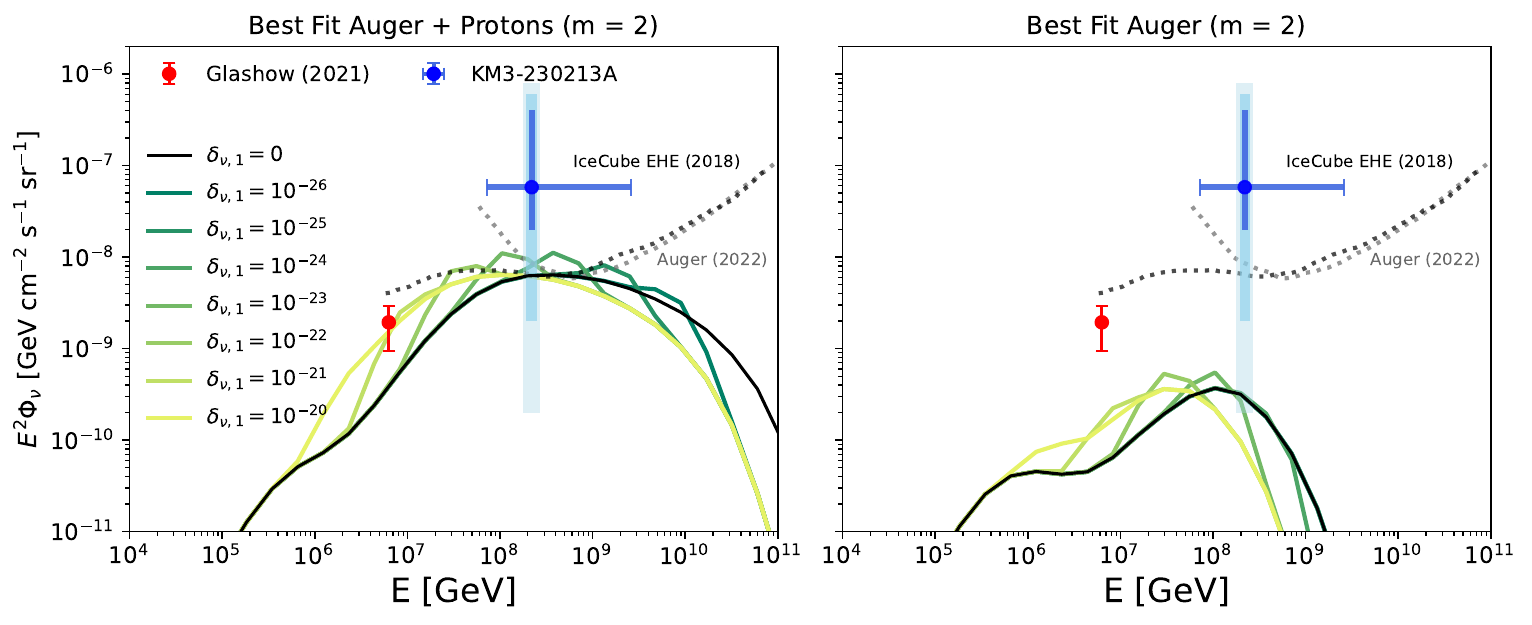}
\caption{Cosmogenic neutrino flux predictions as a function of energy for different LIV coefficients. Left panel shows the best-fit Pierre Auger Observatory model with an additional proton component for $m = 2.0$, while the right panel shows the best-fit Pierre Auger Observatory model~\citep{Aab_2017} without additional protons. Colored curves represent different values of $\delta_{\nu,1}$ from $0$ (standard LI) to $10^{-20} \ \mathrm{eV}^{-1} $. Grey dotted lines show Pierre Auger Observatory 2022~\citep{AbdulHalim:2023SN} constraints, black error bars indicate IceCube EHE 2018~\citep{PhysRevD.98.062003} limits, the red point marks the Glashow 2021~\citep{Aartsen2021} event, and the blue point with error bars shows the KM3-230213A~\citep{2025Natur.638..376K} event. Increasing $\delta_{\nu,1}$ produces a suppression at the highest energies and a corresponding pile-up at lower energies.}
\label{fig:comparison}
\end{figure*}

Figure~\ref{fig:comparison} illustrates the effect of superluminal LIV on the propagated neutrino spectrum for our best-fit UHECR model with source evolution parameter $m = 2.0$. The two panels correspond to the two composition scenarios described in Section~\ref{sec:method}. The left panel shows the proton-enhanced scenario, which combines the best-fit Pierre Auger Observatory parameters from~\citep{Aab_2017} with an additional proton component. The right panel shows the baseline Pierre Auger Observatory-only scenario~\citep{Aab_2017}, without the proton contribution. In both panels, the predicted cosmogenic neutrino flux is compared against observational constraints from the Pierre Auger Observatory, IceCube Extremely High Energy limits, and individual high-energy events including the Glashow resonance~\citep{Aartsen2021} and KM3-230213A~\citep{2025Natur.638..376K}, which serve here as reference points for the LIV constraints.

The comparison between the two panels reveals the role of the assumed primary composition. The proton-enhanced scenario (left panel) yields systematically higher fluxes across all energies, a consequence of the more efficient neutrino production from proton-induced photopion interactions. The pure Pierre Auger Observatory Best Fit composition (right panel) provides a conservative reference in which the standard LI flux is substantially lower. As seen from the green curves ($\delta_{\nu,1} \sim 10^{-20}\ \mathrm{eV}^{-1}$--$10^{-22} \ \mathrm{eV}^{-1}$) in both panels, the LIV flux regeneration effect enhances the neutrino rate in the PeV range regardless of the composition assumption. In the Pierre Auger Observatory scenario, however, the standard LI flux at 6~PeV is negligible, and even with LIV the flux remains well below the level associated with the Glashow resonance event, demonstrating that reaching PeV-scale sensitivity in the conservative baseline requires LIV coefficients large enough to potentially conflict with other constraints. This contrast illustrates that the LIV-induced spectral modification and its observational impact are strongly dependent on the underlying UHECR composition model.

A key feature of the LIV-induced spectral modification concerns the interplay between the energy range of KM3-230213A and the differential upper limits. The standard cosmogenic flux peaks at EeV energies, where constraints from Pierre Auger and IceCube Observatory are most stringent, while the LIV mechanism redistributes flux toward lower energies. The reconstructed energy range of KM3-230213A, spanning roughly 72~PeV to 1~EeV, straddles the region where IceCube upper limits weaken significantly below $\sim 100$~PeV. The LIV pile-up effect suppresses flux in the tightly constrained EeV band and redistributes it precisely into this 50--100~PeV window. This makes KM3-230213A a particularly informative benchmark for probing LIV, since the relevant energy range coincides with the region of maximum spectral modification. Given that the proton-enhanced scenario (left panel) provides the highest baseline flux and therefore the most conservative LIV constraints, the following quantitative discussion focuses on this scenario unless stated otherwise.

%we display the predicted cosmogenic neutrino flux multiplied by $E^2$ as a function of energy, along with observational constraints from the Pierre Auger Observatory 2022 data shown as grey dotted lines, IceCube Extremely High Energy limits from 2018 shown as black error bars, the Glashow resonance event from 2021 marked in red, and the KM3-230213A event highlighted in blue.

%The colored curves illustrate the spectral evolution for LIV coefficients spanning from $\delta_\nu = 10^{-26}$ to $10^{-20}$, alongside the standard Lorentz-invariant baseline (black curve, $\delta_\nu = 0$). 

Increasing $\delta_{\nu,1}$ produces a systematic redistribution of the cosmogenic flux in the proton-enhanced scenario. In the standard LI case (black curve), the predicted flux at $\sim 220$~PeV is approximately $E^2\Phi \sim 3 \times 10^{-9}$~GeV\,cm$^{-2}$\,s$^{-1}$\,sr$^{-1}$, below the single-event flux scale indicated by KM3-230213A ($E^2\Phi \sim 5.8 \times 10^{-8}$~GeV\,cm$^{-2}$\,s$^{-1}$\,sr$^{-1}$). As $\delta_{\nu,1}$ increases, the spectrum does not shift uniformly upward. Instead, a suppression develops at the highest energies where neutrinos become unstable, while a pile-up enhances the flux at lower energies. For moderate coefficients such as $\delta_{\nu,1} = 10^{-24} \ \mathrm{eV}^{-1}$, this enhancement becomes noticeable above $10^7$~GeV. At $\delta_{\nu,1} = 10^{-23} \ \mathrm{eV}^{-1}$, the pile-up raises the flux at 220~PeV to approximately $E^2\Phi \approx 4.8 \times 10^{-9}$~GeV\,cm$^{-2}$\,s$^{-1}$\,sr$^{-1}$, a measurable increase over the LI baseline.

For stronger coefficients, $\delta_{\nu,1} = 10^{-22} \ \mathrm{eV}^{-1}$ or $10^{-21} \ \mathrm{eV}^{-1}$, the flux enhancement in the PeV range approaches and eventually violates the IceCube upper limits at adjacent energies, establishing an upper bound on allowable LIV coefficients. Within the observational constraints currently available, the LIV-induced enhancement in the proton-enhanced scenario is therefore bounded from above. This result does not indicate a failure of the LIV scenario but rather defines the region of parameter space consistent with existing observations, which is precisely the goal of this analysis.

The flux increase in the PeV--EeV range is a direct consequence of energy redistribution through the LIV decay mechanism. High-energy neutrinos become unstable and decay, reinjecting energy at lower energies. This process transfers flux from the highest energies, where the spectrum is suppressed, to the intermediate range, with the secondary neutrinos accumulating just below the decay threshold and producing the characteristic pile-up structure near the energy of KM3-230213A.

\begin{figure}
\centering
\includegraphics[width=\columnwidth]{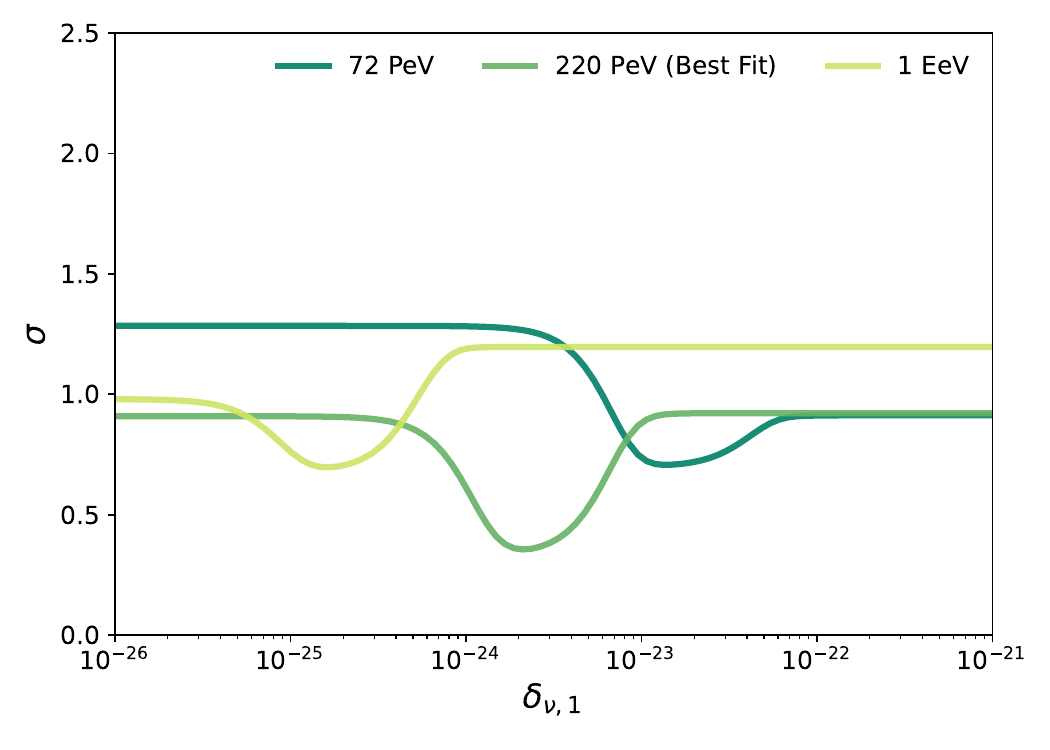}
\caption{Statistical significance $\sigma$ as a function of the LIV coefficient $\delta_\nu$ for three representative energies: 72 PeV, 220 PeV corresponding to the KM3NeT event best-fit energy, and 1 EeV. The non-monotonic behavior at 220 PeV reveals the competition between spectral reshaping and overall flux enhancement, with a minimum near $\delta_{\nu,1} \sim 10^{-23.5} \ \mathrm{eV}^{-1}$. The systematic trends at 72 PeV (decreasing) and 1 EeV (increasing) demonstrate the energy-dependent nature of LIV effects.}
\label{fig:sigma}
\end{figure}

To quantify how LIV modifies the statistical weight of a single event detection, we compute the significance $\sigma$ as a function of $\delta_{\nu,1}$ at three representative energies spanning the reconstructed energy range of KM3-230213A: 72~PeV, 220~PeV (best-fit), and 1~EeV. These energies serve as markers to trace the energy-dependent response of the spectrum to LIV and are evaluated using the proton-enhanced scenario. Figure~\ref{fig:sigma} shows the result.

At 220~PeV, the standard LI baseline yields $\sigma \sim 0.9$. As $\delta_{\nu,1}$ increases, the significance first drops to a minimum of $\sigma \sim 0.35$ near $\delta_{\nu,1} \sim 10^{-23.6} \ \mathrm{eV}^{-1}$, reflecting a region where spectral reshaping locally suppresses the flux at that specific energy, before returning to a plateau of $\sigma \sim 0.9$ at higher coefficients. At 72~PeV, the significance starts at $\sigma \sim 1.3$, displays a localized dip to $\sigma \sim 0.7$ near $\delta_{\nu,1} \sim 10^{-22.8} \ \mathrm{eV}^{-1}$, and then reaches a plateau of $\sigma \sim 0.9$ for $\delta_{\nu,1} > 10^{-22} \ \mathrm{eV}^{-1}$, consistent with superluminal LIV preferentially shifting flux toward higher energies within the PeV band. At 1~EeV, the significance starts at $\sigma \sim 1.0$, displays a localized dip to $\sigma \sim 0.7$ near $\delta_{\nu,1} \sim 10^{-25}$, and then rises to a flat plateau of $\sigma \sim 1.2$ above $\delta_{\nu,1} \sim 10^{-24} \ \mathrm{eV}^{-1}$. For $\delta_{\nu,1} > 10^{-22.5} \ \mathrm{eV}^{-1}$, complete saturation occurs and all three curves become independent of further variation in the LIV parameter.

\subsection{Testing the Observational Viability}

\begin{figure}
\centering
\includegraphics[width=\columnwidth]{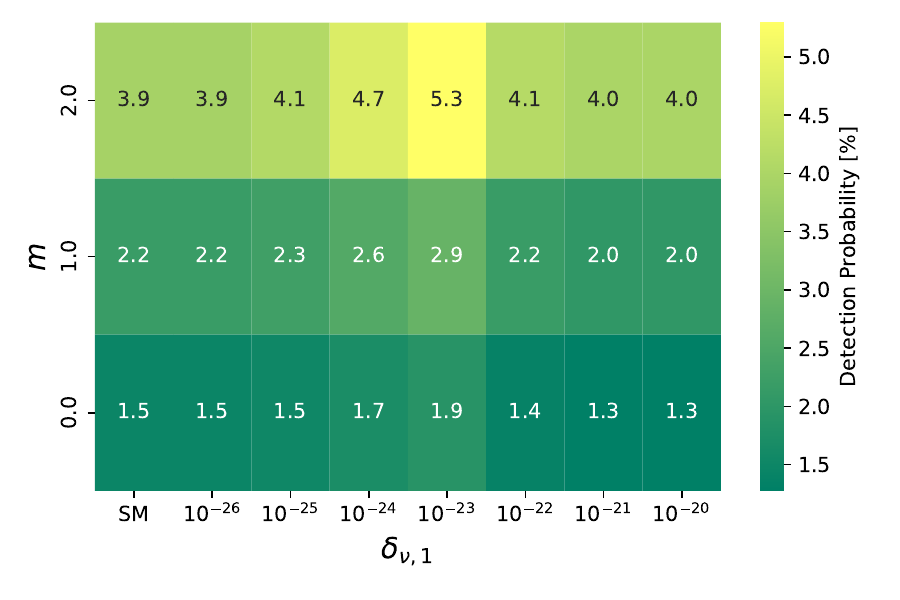}
\caption{Detection probability of at least one cosmogenic neutrino event in the KM3NeT ARCA detector, spanning the parameter space of the source evolution index $m$, the LIV parameter $\delta_{\nu,1}$ and considering the Best Fit Pierre Auger Observatory + Protons scenario~\citep{Aab_2017}. The Standard Model (SM) prediction without LIV ($\delta_{\nu,1} = 0$) is shown for comparison.}
\label{fig:heatmap_prob}
\end{figure}

To evaluate the observational viability of our LIV scenarios across the full parameter space, we compute the expected number of neutrino events and the corresponding detection probability for each configuration of the source evolution index $m$ and $\delta_{\nu,1}$, using the proton-enhanced scenario throughout. Following the approach of the KM3NeT Collaboration~\citep{2025Natur.638..376K}, the KM3NeT/ARCA all-sky exposure is defined as:
\begin{equation}
\mathcal{E}^{\text{KM3NeT}}(E) = 4\pi \times T_{\text{KM3NeT}} \times A_{\text{eff}}^{\text{KM3NeT}}(E)
\end{equation}
in which $T_{\text{KM3NeT}} = 335$~days and $A_{\text{eff}}^{\text{KM3NeT}}(E)$ is the sky-averaged effective area for the bright track selection, averaged between neutrinos and antineutrinos.

Assuming a standard oscillation-averaged flavour equipartition at Earth ($\nu_e:\nu_\mu:\nu_\tau \approx 1:1:1$), the per-flavour $\nu+\bar{\nu}$ flux $\Phi(E)$ is obtained by taking one-third of the total all-flavour cosmogenic flux generated by our simulations. The expected number of events for this per-flavour all-sky flux is then
\begin{equation}
n_{\text{expected}}^{\text{KM3NeT}} = \int_{E_{\text{min}}}^{E_{\text{max}}} \mathcal{E}^{\text{KM3NeT}}(E) \times \Phi(E) \, dE,
\end{equation}
where the integration bounds are fixed to the central 90\% neutrino energy range associated with KM3-230213A, spanning from $E_{\text{min}} = 7.24 \times 10^7$~GeV to $E_{\text{max}} = 2.57 \times 10^9$~GeV. The probability of detecting at least one event is then computed via a Poisson estimator,
\begin{equation}
P(\ge 1) = 1 - e^{-n_{\text{expected}}^{\text{KM3NeT}}}.
\end{equation}

Figure~\ref{fig:heatmap_prob} presents the detection probability across the full range of source evolution indices $m$ and LIV parameters $\delta_{\nu,1}$ for the proton-enhanced scenario. While the Standard Model (SM) yields relatively low detection probabilities across the considered source evolution indices ($m \geq 0.0$), the introduction of superluminal LIV systematically enhances the likelihood of observation. The region $\delta_{\nu,1} \sim 10^{-24}$--$10^{-22}$ emerges as a robust band in which the detection probability is maximised while remaining consistent with the energy-dependent constraints discussed above.

Together, the significance and detection probability analyses identify a preferred range of LIV coefficients that is observationally viable. Values in the range $10^{-24} \ \mathrm{eV}^{-1} \lesssim \delta_{\nu,1} \lesssim 10^{-22} \ \mathrm{eV}^{-1}$ maximise the expected detection rate in the proton-enhanced scenario while avoiding overproduction of neutrinos in bands where IceCube non-detections place tight constraints. This range corresponds to LIV energy scales of $E_{\rm LIV} \sim (|\delta_{\nu,1}|)^{-1} \sim 10^{22}$--$10^{24}$~eV, approximately five to six orders of magnitude below the Planck scale, but well above the energies accessible in terrestrial experiments.

\section{Summary and Conclusions}
\label{sec:conclusions}

In this work, we have characterised superluminal LIV in the neutrino sector through its imprint on cosmogenic neutrino fluxes, using the KM3-230213A event as a benchmark to probe the accessible parameter space. Our analysis demonstrates that superluminal LIV induces a spectral suppression at the highest energies accompanied by a pile-up enhancement at PeV energies. This secondary flux enhancement is most pronounced for LIV coefficients in the range $10^{-24} \ \mathrm{eV}^{-1} \lesssim \delta_{\nu,1} \lesssim 10^{-22} \ \mathrm{eV}^{-1}$, which represents the observationally viable window constrained from above by IceCube non-detections and characterised from below by the sensitivity of the KM3NeT observation.

The corresponding LIV energy scale, $E_{\text{LIV}} \sim 10^{22}$--$10^{24}$~eV, lies approximately five to six orders of magnitude below the Planck scale. While such suppression is not predicted in the simplest Planck-scale scenarios, it remains consistent with several quantum-gravity frameworks, including loop quantum gravity and certain string theory compactifications, where LIV effects can appear with additional suppression factors~\citep{1998Natur.393..763A, 2006AnPhy.321..150J}. Importantly, this constraint applies specifically to the neutrino sector and does not conflict with existing, tighter bounds on LIV in the hadronic and photonic sectors derived from UHECR spectra and $\gamma$-ray transparency~\citep{Lang2022, 2020Symm...12.1232M}.

A key methodological strength of this approach is the analytic treatment of neutrino propagation. Unlike cosmic rays or high-energy photons, neutrinos do not undergo further interactions after production, allowing LIV effects to be incorporated via a direct modification of the dispersion relation without cascade simulations. This makes cosmogenic neutrinos a particularly clean probe of fundamental physics at energy scales inaccessible to terrestrial experiments~\citep{2009APh....31..220S}. Several testable predictions follow from our results. The expected spectral hardening is most pronounced above $\sim 100$~PeV and could be probed with increased statistics from \texttt{IceCube-Gen2}~\citep{2021JPhG...48f0501A}. If LIV coefficients are flavour-dependent, deviations from the standard oscillation-averaged $\nu_e:\nu_\mu:\nu_\tau \approx 1:1:1$ ratio at Earth could emerge~\citep{2015PhRvL.115p1302B}. Upcoming facilities such as \texttt{GRAND} and \texttt{POEMMA}, sensitive to EeV neutrinos, will provide complementary tests at the highest energies where LIV effects are strongest~\citep{2020SCPMA..6319501A, 2021JCAP...06..007P}.

Systematic uncertainties in UHECR source parameters, particularly composition, injection spectrum, and redshift evolution, propagate to the neutrino flux predictions and broaden the allowed LIV parameter space~\citep{2008JCAP...10..033A, 2014JCAP...10..020A}. Future multi-messenger observations and improved statistics will reduce these uncertainties and sharpen the LIV constraints.

In conclusion, this study demonstrates that ultra-high-energy cosmogenic neutrinos provide a sensitive and clean probe of Planck-scale physics in the neutrino sector. The KM3-230213A event, used here as a reference benchmark rather than a target for explanation, yields a variation of approximately $0.5\sigma$ and thus lacks the statistical significance to formally constrain the parameter space, it successfully highlights the potential of this approach. Specifically, our analysis demonstrates the methodology sensitivity to superluminal LIV in the range $10^{-24} \lesssim \delta_{\nu,1} \lesssim 10^{-22}$ within realistic UHECR models. As data from upcoming neutrino telescopes accumulate with increased event statistics, cosmogenic neutrinos will provide increasingly stringent tests of Lorentz invariance, potentially revealing signatures of quantum gravity in the neutrino sector.

\section*{Acknowledgements}

This study was financed in part by the Coordenação de Aperfeiçoamento de Pessoal de Nível Superior-Brasil (CAPES) - Finance Code 001. R.S and R.C.A. acknowledge the financial support from the NAPI “Fenômenos Extremos do Universo' of the Fundação de Apoio à Ciência, Tecnologia e Inovação do Paraná. R.C.A. research is supported by CNPq (308859/2025-1) and (4000045/2023-0), Araucária Foundation (698/2022) and (721/2022), and FAPESP (2021/01089-1). The authors express their sincere gratitude to the National Laboratory for Scientific Computing (LNCC/MCTI, Brazil) for providing HPC resources through the SDumont supercomputer and LIneA (Interinstitutional Laboratory Association for e-Astronomy), which significantly supported the computational aspects of this research, which have contributed to the research results reported in this paper. URL: https://sdumont.lncc.br and https://www.linea.org.br/.

%%%%%%%%%%%%%%%%%%%%%%%%%%%%%%%%%%%%%%%%%%%%%%%%%%
\section*{Data Availability}

The data underlying this article were generated using the open-source simulation framework \textit{CRPropa 3.2}, which is available at \url{https://crpropa.github.io/CRPropa3/}. The specific input files and configuration used to generate the cosmogenic neutrino fluxes and LIV modifications presented in this work will be shared on reasonable request to the corresponding author.



\bibliographystyle{mnras}
\bibliography{example} % if your bibtex file is called example.bib

%\begin{thebibliography}{99}
%\end{thebibliography}

%%%%%%%%%%%%%%%%%%%%%%%%%%%%%%%%%%%%%%%%%%%%%%%%%%

%%%%%%%%%%%%%%%%% APPENDICES %%%%%%%%%%%%%%%%%%%%%

%\appendix



%%%%%%%%%%%%%%%%%%%%%%%%%%%%%%%%%%%%%%%%%%%%%%%%%%

% Don't change these lines
\bsp	% typesetting comment
\label{lastpage}
\end{document}